# Milliwatt terahertz harmonic generation from topological insulator metamaterials

Klaas-Jan Tielrooij[1,2,*], Alessandro Principi[3], David Saleta Reig[1], Alexander Block[1], Sebin Varghese[1], Steffen Schreyeck[4], Karl Brunner[4], Grzegorz Karczewski[4,5], Igor Ilyakov[6], Oleksiy Ponomaryov[6], Thales V. A. G. de Oliveira[6], Min Chen[6], Jan-Christoph Deinert[6], Carmen Gomez Carbonell[1], Sergio O. Valenzuela[1,7], Laurens W. Molenkamp[4,8], Tobias Kiessling[4], Georgy V. Astakhov[6,†], and Sergey Kovalev[6,‡]

[1] *Catalan Institute of Nanoscience and Nanotechnology (ICN2), BIST and CSIC, Campus UAB, Bellaterra (Barcelona), 08193, Spain*
[2] *Department of Applied Physics, TU Eindhoven, Den Dolech 2, 5612 AZ Eindhoven, The Netherlands*
[3] *School of Physics and Astronomy, University of Manchester, M13 9PL, Manchester, UK*
[4] *Physikalisches Institut (EP3), Universität Würzburg, Am Hubland, 97074 Würzburg, Germany*
[5] *Institute of Physics, Polish Academy of Science, Al. Lotnikow 32/46, PL-02668 Warsaw, Poland*
[6] *Helmholtz-Zentrum Dresden-Rossendorf, Bautzner Landstr. 400, 01328 Dresden, Germany*
[7] *ICREA, Pg. Lluís Companys 23, 08010 Barcelona, Spain*
[8] *Institute for Topological Insulators, Am Hubland, D-97074 Würzburg, Germany*

[*]E-mail: klaas.tielrooij@icn2.cat
[†]E-mail: g.astakhov@hzdr.de
[‡]E-mail: s.kovalev@hzdr.de

**Abstract**

Achieving efficient, high-power harmonic generation in the terahertz spectral domain has technological applications, for example in sixth generation (6G) communication networks. Massless Dirac fermions possess extremely large terahertz nonlinear susceptibilities and harmonic conversion efficiencies. However, the observed maximum generated harmonic power is limited, because of saturation effects at increasing incident powers, as shown recently for graphene. Here, we demonstrate room-temperature terahertz harmonic generation in a $Bi_2Se_3$ topological insulator and topological-insulator-grating metamaterial structures with surface-selective terahertz field enhancement. We obtain a third-harmonic power approaching the milliwatt range for an incident power of 75 mW – an improvement by two orders of magnitude compared to a benchmarked graphene sample. We establish a framework in which this exceptional performance is the result of thermodynamic harmonic generation by the massless topological surface states, benefiting from ultrafast dissipation of electronic heat via surface-bulk Coulomb interactions. These results are an important step towards on-chip terahertz (opto)electronic applications.

**Introduction**

One of the most canonical manifestations of nonlinear interaction of electromagnetic waves with matter is harmonic generation, where the energy of generated photons is a multiple of the energy of incident photons. Until a few years ago, the most efficient way of generating harmonics in the terahertz (THz) regime was using two-dimensional quantum well systems [1–3]. During the past few years several other material systems, including $SrTiO_3$ [4] and superconductors [5] have shown strong THz nonlinearities. In particular, it has been demonstrated that materials with massless Dirac fermions – charge carriers with a linear energy-momentum dispersion relation – can lead to efficient THz harmonic generation. This has been shown with the prototypical massless Dirac-fermion material graphene [6], with topological insulators (TIs) [7, 8], and with Dirac semi-metals [9, 10]. Currently, the THz nonlinear photonics "toolbox" includes the possibility to electrically tune harmonic generation [11], and the ability to employ ultrathin metamaterials that generate an increased harmonic power by orders of magnitude, as a result of field enhancement in a metallic grating located nearby the nonlinear Dirac-fermion system [12]. This has resulted in a record-high third-order susceptibility $\chi^{(3)}$ above $10^{-8}$ m² V⁻² in the THz regime for a grating-graphene metamaterial [12], which is more than three orders of magnitude larger than the highest value obtained with quantum wells [1–3]. In order to fully exploit these ultrahigh THz nonlinearities towards practical technologies, there are fundamental and technological obstacles that need to be overcome. Important fundamental questions pertain to the mechanism that leads to THz harmonic generation in massless Dirac-fermion systems, and to whether THz harmonic generation in TIs originates mainly from the surface states or from bulk carriers without linear dispersion. For technological applications, it is necessary to significantly increase the amount of generated harmonic power.

**Results**

Here, we address these issues by comparing THz third-harmonic generation in a 102 nm thick $Bi_2Se_3$ TI sample with the response from monolayer graphene serving as benchmark sample (see Fig. 1). The TI sample is optimized towards low bulk carrier density (see Methods). In Figure S1 we also show results for 50 nm thick $Bi_2Te_3$. We chose these material systems because they combine the presence of massless Dirac fermions with ultrashort cooling times [8], which could be beneficial for harmonic generation. On each sample, we fabricated a region with a metallic grating (see Fig. 1a–b). The reason for adding a metallic grating is that it gives rise to field enhancement inside the gaps, while not sacrificing too much active area. This leads to increased harmonic generation, as shown in Ref. [12]. The metallic grating consists of metal strips with width $w_{\text{metal}}$, separated by a gap with width $w_{\text{gap}}$. The graphene sample has a fill factor of $F = \frac{w_{\text{metal}}}{w_{\text{metal}} + w_{\text{gap}}} \approx 90\%$ and the TI sample has $F \approx 85\%$.

We study the generation of THz third harmonics, where generated photons have triple the energy of incident photons, using narrowband THz pulses with a fundamental frequency $f = 0.5$ THz, and a duration of around 10 ps, generated at the superradiant THz facility TELBE located at Helmholtz Zentrum Dresden Rossendorf [13] (see Methods for details). We focus this light onto a sample and measure the transmitted waveform using electro-optic sampling (EOS) with 100 fs laser sampling pulses and a 2 mm thick ZnTe nonlinear crystal. This provides information on the temporal evolution of the electric field. Fourier transformation of this time-domain signal then provides the spectral components of the transmitted light, including signal at the third harmonic, $3f = 1.5$ THz. We point out that in these centrosymmetric materials even harmonics are forbidden.

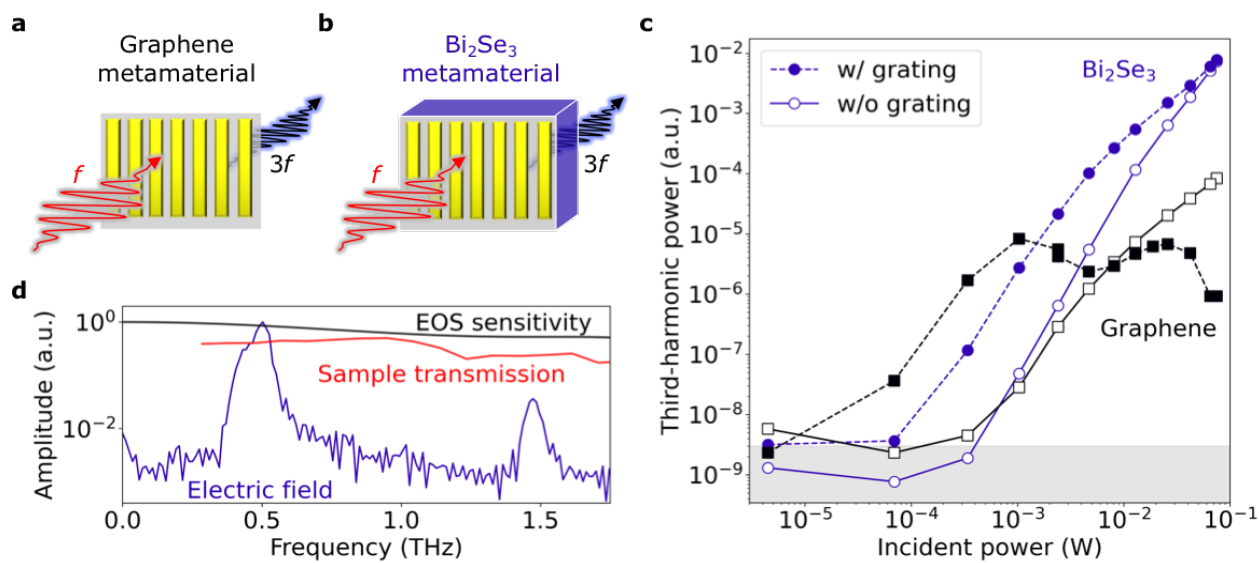

**Fig. 1: THz harmonic generation in Bi$_2$Se$_3$ *vs.* graphene. a–b)** Schematic experimental layout of third-harmonic generation studies in graphene **(a)** and Bi$_2$Se$_3$ **(b)** metamaterials. **c)** The THz third-harmonic power is enhanced by the metallic gratings for both graphene (black squares and lines) and Bi$_2$Se$_3$ (blue circles and lines). However, the power saturates strongly for graphene, and therefore eventually Bi$_2$Se$_3$ outcompetes graphene by around two orders of magnitude. The largest pump intensity corresponds to an incident peak field strength of 300 kV cm$^{-1}$. **d)** Transmitted spectral amplitude (normalized) after the Bi$_2$Se$_3$ metamaterial sample (blue line), with a $3f$ signal that is ~4% of the fundamental signal at $f$. Both the sample transmission (red line) and electrooptic sampling (EOS) sensitivity of the 2 mm thick ZnTe detection crystal (black line) are lower at $3f$ than at $f$. Taking into account the lower EOS sensitivity, the field-conversion efficiency is 8%, corresponding to 0.5 mW of generated harmonic power.

In Fig. 1c we show the power of the third-harmonic signal for increasing incident fundamental power for regions with and without metal grating. Both graphene and TI samples show enhancement due to the metal grating, similar to Ref. [12]. However, the behavior at high incident power is remarkably different: Whereas the signal from graphene completely saturates at high pump intensities, the signal from the TI keeps increasing, eventually surpassing the signal produced in graphene by orders of magnitude. We quantify the maximum conversion efficiency for the TI sample in Fig. 1d, and we find that the third-harmonic signal is ~25 times smaller than the fundamental. Taking into account the reduced EOS sensitivity at 3*f*, we infer a field conversion efficiency of ~8%. With an input power of $P_f = 75$ mW, this means that the TI sample is generating around $P_{3f} = 0.5$ mW of third-harmonic power.

In order to understand why saturation is much stronger in graphene than in the TI, we examine the electric field signals in the time domain. Figure 2a shows the temporal evolution of the electric field for a typical incident waveform at 0.5 THz. For graphene using the weakest incident field strength of 9 kV cm$^{-1}$, and filtering the signal at the fundamental frequency, the envelope of the transmitted pulse is similar to that of the incident waveform (see Fig. 2b). However, the dynamics change dramatically upon increasing the incident field strength: the signal in the time domain becomes stretched, and even has some long-term oscillatory behavior (see Fig. 2b). We ascribe this to heat accumulation effects: when the electronic system is heated significantly, the cooling to acoustic phonons becomes slower, and optical phonons can reheat the electronic system, resulting in slower cooling [14]. This leads to heat accumulation and to such high electron temperatures that THz absorption becomes nearly zero. Such saturable THz absorption in graphene was observed in Ref. [15] and can be ascribed to several phenomena, as reviewed in Ref. [16]. During periods of negligible THz absorption, we expect negligible THz harmonic generation. Only when the temperature has dropped some picoseconds later can the system absorb incident THz light and generate THz harmonic signal. The delicate balance between THz-electronic heating and cooling through optical and acoustic phonons during the interaction with the incident multicycle THz pulse thus explains the dips in the envelope of the signal, and the strong decrease in overall harmonic generation efficiency. When we examine the envelopes of the signal for $Bi_2Se_3$ (see Fig. 2c), we do not observe any distortions, which means that, in contrast to graphene, the TI sample does not suffer from any detrimental heat accumulation effects (see Fig. 2d–e). As a result, the TI exhibits much weaker saturation, leading to a significantly higher harmonic conversion efficiency at higher incident powers and thus a larger harmonic output power. We also note that within the framework of thermodynamic harmonic generation, first discussed for massless Dirac carriers in graphene [6, 17], fast electron cooling is beneficial for efficient harmonic generation. Within this framework, continuous electron heating-cooling dynamics occur during the interaction with an incident multi-cycle THz pulse. These dynamics lead to a modulating electron temperature, which corresponds to a modulated THz absorption, due to the electron-temperature dependent THz conductivity [16, 18]. Faster cooling leads to a larger modulation of the electron temperature and therefore a larger harmonic generation efficiency.

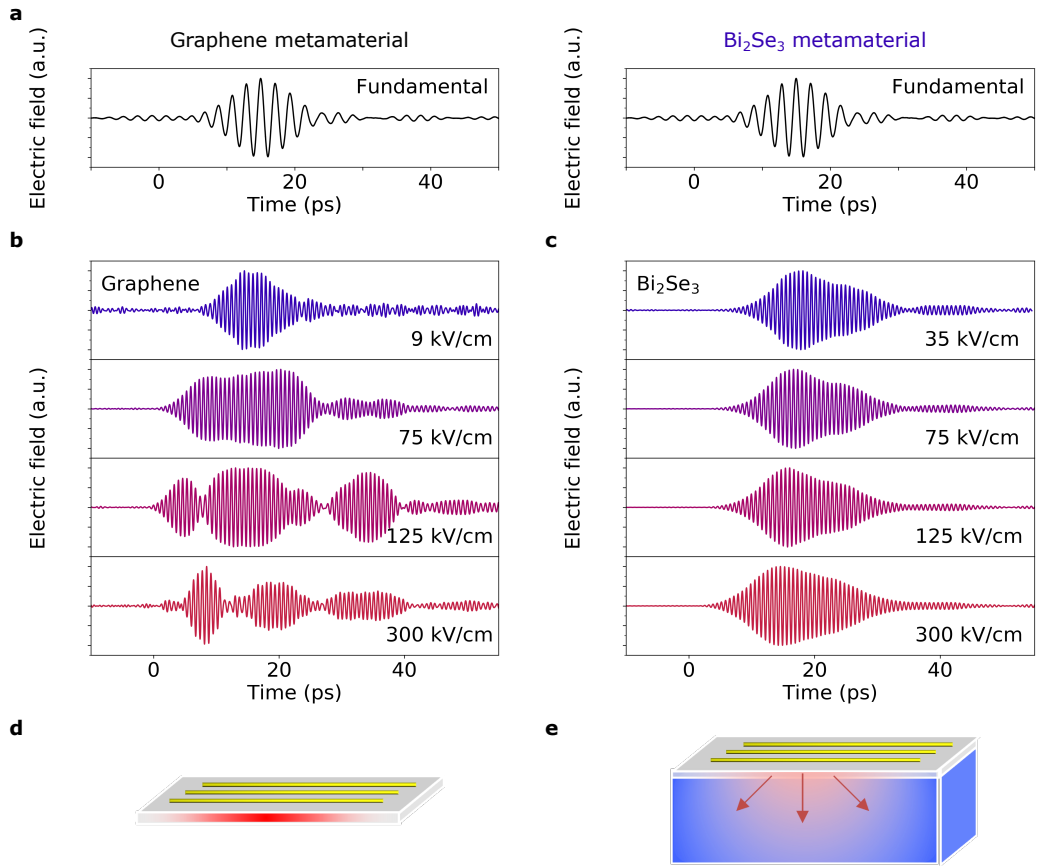

**Fig. 2: Time-domain evolution of harmonic generation. a)** Time-domain dynamics (normalized) of a typical fundamental pulse, as measured through electro-optic sampling. **b–c)** Time-domain dynamics (normalized) of generated third-harmonic signal for graphene **(b)** and $Bi_2Se_3$ **(c)**, for four different incident peak field strengths. Both samples are measured in the region with a metal grating for field enhancement. Graphene displays complex dynamics because of the interplay between electronic and phononic heat accumulation and heat-induced (near-)complete THz transparency, which leads to saturation of harmonic generation. Due to the efficient surface-to-bulk dissipation of electronic heat in $Bi_2Se_3$, no heat accumulation occurs, and therefore the envelope of the time-domain signal basically follows that of the incident, fundamental waveform. **d–e)** Schematic illustration of heat accumulation in graphene **(d)** and efficient surface-to-bulk dissipation in the TI **(e)**.

**Discussion**

We ascribe the more efficient heat dissipation in TIs to highly efficient Coulomb scattering between surface-state charges and bulk charges – a process similar to the interlayer dissipation occurring in multilayer graphene [19]. We calculate the cooling time that corresponds to this surface-bulk interaction mechanism using a Boltzmann equation approach (see Ref. [20] for details). The employed electron-electron collision integral conserves particle number in each of the separate systems (surface and bulk), while allowing for energy exchange. We perform the calculations with two sets of input parameters, where the Fermi velocity and bulk effective mass are $10^6$ m s$^{-1}$ and $0.14 m_e$, or $0.4 \cdot 10^6$ m s$^{-1}$ and $0.21 m_e$, where $m_e$ is the electron mass. For the bandgap we take 200 meV, and for the permittivity we take $\varepsilon = 10$. The Fermi energy is in the bandgap with a carrier density of $10^{12}$ cm$^{-2}$ for the surface states. With these parameters, which are based on Refs. [7, 21], we find a cooling time for surface-state carriers around 300 fs for an electron temperature between 500 and 2000 K (see Fig. 3a). This is in good agreement with a recent THz-pump optical-probe experiment [8]. Importantly, surface-to-bulk "Coulomb cooling" is faster than electron-phonon coupling in the bulk, which is typically found to occur on a timescale of a few picoseconds [7, 22 23]. We point out that this mechanism will likely be less efficient when the Fermi energy is located in the bulk of the TI. The picture of increased harmonic conversion efficiency due to efficient surface-to-bulk electronic heat dissipation is thus consistent with the framework of thermodynamic harmonic generation, specifically enabled by the massless Dirac surface states, which exhibit ultrafast cooling.

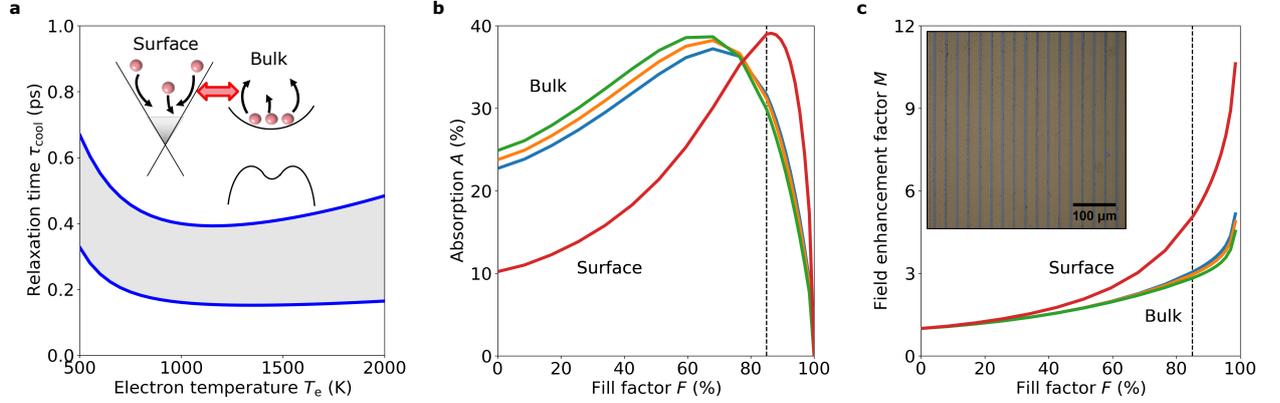

**Fig. 3: Surface-bulk interplay. a)** Calculation results for the cooling time of surface-state carriers by dissipation of electronic heat to bulk carriers, for two sets of input parameters. The inset is a schematic of the microscopic mechanism where surface-state carriers lose energy by Coulomb interaction with bulk carriers, without surface-to-bulk charge transfer. The obtained timescale of this "Coulomb cooling" is in agreement with a recent study of ultrafast carrier cooling of TI surface states [8]. **b–c)** RCWA simulation results for total THz absorption **(b)** and field enhancement **(c)** for a 3D simulation using the properties of bulk carriers (blue, orange, green lines; corresponding to permittivities of 10, 180, 360, respectively) and a 2D simulation using the properties of the surface-state carriers (red lines). The grating effect is much larger for the surface-state carriers, where significant field enhancement occurs. For our sample with fill factor $F = 85\%$ (see microscope image in the inset of panel **c**, where the brighter stripes are the metal grating), we find five-fold field enhancement.

We provide further evidence that the harmonic signal is generated predominantly from the massless Dirac surface states in TIs by simulating the effect of the metallic grating on the surface and on the bulk, using rigorous coupled wave analysis (RCWA) simulations [24] (see Methods for details). In the former case, we simulate the TI as a purely 2D conducting sheet with the electronic properties of the surface states, and in the latter case, we use a 3D conducting volume with the electronic properties of the bulk. The results in Fig. 3b show that the absorption in the bulk is already quite high, and therefore cannot be increased much by the metal grating. For fill factors above 60%, the overall absorption even decreases, because of the reduced active area. For the surface-state carriers, the intrinsic absorption is lower, because the carrier density is lower, and therefore the absorption can be enhanced substantially by the grating. Figure 3c shows the field enhancement factor $M_{\text{field}}$, which is significantly higher for the surface-state carriers ($M_{\text{field}} \approx 5$) than for the bulk carriers ($M_{\text{field}} \approx 2.6$).

Since the metallic gratings lead to surface-selective enhancement, this enables us to distinguish between surface-generated and bulk-generated harmonic signals by comparing the experimentally observed harmonic enhancement with those calculated for harmonics from the bulk and from the surface. The power enhancement factor for the third harmonic in the perturbative regime is given by $M_{3f} = \frac{P_{3f,\text{grating}}}{P_{3f,\text{no grating}}} = M_{\text{field}}^6 \cdot (1 - F)$ (see Ref. [12]). The $(1 - F)$ term represents the effect of the reduced active area due to the metal grating, and the 6th power comes from the scaling between the generated third-harmonic power and the incident field strength. For intermediate incident powers, around 1–10 mW, we estimate that this scaling is closer to the 4th power, as the third-harmonic power scales roughly quadratically with incident power (see Fig. 1c). Therefore, we expect the harmonic power enhancement factor $M_{3f}$ to be $M_{\text{field}}^4 \cdot (1 - F)$, which gives ~7 for the bulk and ~100 for the surface states. The experimentally observed harmonic enhancement factor for $Bi_2Se_3$ (see Fig. 1c) and $Bi_2Te_3$ (see Figure S1) at an incident power around 1 mW is $M_{3f} = 100\text{–}300$. This is therefore in much better agreement with harmonic generation from surface states than from bulk states, which is also supported by the conclusion of Ref. [7]. Moreover, it is clear that with a larger fill factor, even larger third-harmonic enhancement, and possibly output intensities, will be possible.

We note that some signatures of saturation do occur for the TI-based samples, and in particular for the TI-grating metamaterial (see Fig. 1c). This also explains why the effect of the metallic grating is quite weak for the highest incident power: the grating works best in the perturbative regime, where the third-harmonic power scales with the sixth power of the field enhancement factor. There are a number of reasons for the observed non-perturbative behavior. First of all, the electron temperature of two-dimensional, massless Dirac fermions intrinsically scales sub-linearly with the incident power [12]. Furthermore, it is possible that the bulk of the TI becomes increasingly metallic for increasing incident powers, leading to absorption of THz harmonic signal that is generated at the front surface. Importantly, it remains to be seen if even larger third-harmonic output powers will be achieved with or without the grating. Nevertheless, it is clear that we have not yet seen the limit of what is possible, as the generated harmonic power was still increasing up to the largest incident power that was available.

In conclusion, we have demonstrated that topological insulators generate efficient and high-power THz third-harmonic radiation, with a field conversion efficiency of ~8% and a power approaching the mW level. We interpret the results within the framework of harmonic generation through the thermodynamic nonlinearity mechanism of massless Dirac fermions, as introduced in Refs. [6, 15]. We attribute the saturation effects in graphene to heat accumulation in both electronic and phononic systems. This heat accumulation does not occur in the TI, because of ultrafast dissipation of electronic heat through "Coulomb cooling" of surface states to the bulk. Therefore, the THG power generated by the TI is surpassing that from graphene by orders of magnitude for high incident powers. Combining the TI with a metal grating results in a harmonic power enhancement of two orders of magnitude for low to intermediate input powers. This is consistent with harmonic generation by surface states, rather than the bulk, as the harmonic enhancement by the grating is surface-selective. The effectiveness of the grating is of technological importance, as it enables reaching similar output intensities for smaller input fields, and can be optimized further. The results obtained here for the prototypical TIs $Bi_2Se_3$ and $Bi_2Te_3$ are likely representative of the broad class of TI materials with massless Dirac carriers. Our findings thus pave the way for on-chip, technologically-friendly high-power THz photonics.

**Materials and Methods**

*Sample fabrication, Bi$_2$Se$_3$* – The Bi$_2$Se$_3$ sample was grown by molecular beam epitaxy on a semi-insulating two-inch InP(111)B wafer. As substrate we choose the nearly perfect lattice matched InP(111)B with a surface roughness root mean square of about 2 nm to achieve a high-quality film with completely suppressed twinning, which is otherwise a ubiquitous defect in Bi$_2$Se$_3$ thin films [25]. The sample is grown at a substrate temperature of 300 °C by co-deposition of 6N Bi and 6N Se by effusion cells at a growth rate of about 1 nm per second in Se-rich conditions [25]. After growth, the sample is cooled to 140 °C in Se flux to prevent Se out-diffusion at elevated temperatures. The 102 nm thick Bi$_2$Se$_3$ film was characterized by X-ray diffraction, transmission electron microscopy, atomic force microscopy, X-ray reflectometry, and magneto-optics, as shown in Refs. [25, 26]. Hall measurements at a temperature of 4 K showed an n-type carrier density of $9 \cdot 10^{17}$ cm$^{-3}$ and a carrier mobility of 2000 cm$^2$ V$^{-1}$ s$^{-1}$ (assuming a one-carrier model) [25]. This sample has been exposed to air since 2012 and has not shown any indication of degradation, which is promising for technological applications.

*Sample fabrication, Bi$_2$Te$_3$* – Untwinned Bi$_2$Te$_3$ films were grown on BaF$_2$ (111) substrates (1 × 1 cm$^2$) by molecular beam epitaxy under ultra-high vacuum (UHV) conditions in a dual-chamber reactor (Createc) with a base pressure of $2 \cdot 10^{-10}$ Torr [27]. The growth was carried out at 400 °C by co-evaporation of Bi and Te using solid cracker cells with a rate of 0.3 QL min$^{-1}$, while monitoring surface reconstruction in-situ by 15-KeV reflection high-energy electron diffraction (RHEED). Calibration was implemented using RHEED oscillations and X-ray reflectivity. Prior to growth, the Bi and Te fluxes were stabilized using a quartz-crystal microbalance. The cell/cracker temperatures were 670/1100 °C for Bi and 261/600 °C for Te. The Te/Bi flux ratio was kept around 2, slightly larger than the stoichiometric ratio of 1.5, to compensate for Te desorption during growth. The final film thickness was 50 nm. After growth and to prevent the Bi$_2$Te$_3$ from oxidizing, the sample was transferred to a second UHV chamber through a UHV channel, where a thin capping layer, consisting of 3 nm of Al, was deposited and let to oxidize in air. Finally, the sample was taken to an atomic layer deposition reactor (Cambridge NanoTech) where a final layer of 10 nm Al$_2$O$_3$ was grown for subsequent fabrication steps.

*Sample fabrication, gratings* – Gold gratings with areas of 1.5–2.5 mm$^2$ and fill factors $F$ up to 85% were fabricated onto Al$_2$O$_3$-capped TI samples by direct write laser lithography (Kloé Dilase 250). For Bi$_2$Se$_3$ samples, positive photoresist AZ5214E (MicroChemicals GmbH) was spin-coated on the TI and baked at 100 °C for 1 minute. The photoresist was then exposed with optimal dose and developed to reveal the TI surface under the exposed regions. For Bi$_2$Te$_3$ samples, the gap regions were exposed and the pattern was inverted by post-exposure baking at 120 °C for 2 minutes followed by ultraviolet exposure (Kloé KLUB 3) with a power density of 35 mW cm$^{-2}$ for 10 seconds. Inductively-coupled plasma - reactive ion etching (Oxford PlasmaPro 100 Cobra) was used to reduce the presence of residues with a mild oxygen plasma prior to e-beam evaporation (AJA Orion) of Ti/Au (5/50 nm). Last, the lift-off was carried in acetone ($>$ 2 h) aided with mild ultrasounds (10 seconds).

*THz measurements* – For THz harmonic measurements we used the superradiant accelerator-based source TELBE. TELBE was tuned to generate narrowband THz radiation with 0.5 THz central frequency and about 10 ps pulse duration. To block any radiation outside the pump pulse bandwidth two bandpass filters with 0.5 THz central frequency and 20% bandwidth were inserted into the beam path before the investigated sample. Using off-axis parabolic mirrors the THz pump pulse was focused onto the sample surface and had about 400 µm full-width-at-half-maximum spot size. The Spiricon PyrocamIII was used to image the THz beam profile on the sample. The THz radiation transmitted through the samples was refocused onto a 2 mm thick ZnTe crystal and standard THz time-domain spectroscopy was used to measure the temporal profile of the radiation. In order not to saturate the electro-optic sampling crystal with the fundamental radiation, we either used 1.5 THz bandpass filters after the sample (for Fig. 1c and Fig. 2), or attenuated the fundamental and third-harmonic radiation with a pair of wire grid polarizers, having the same attenuation for fundamental and harmonic frequencies (Fig. 1d). For the electro-optical sampling, 100 fs laser pulses were used for probing. The timing jitter between the TELBE source and the laser probe pulses was on the order of a few tens of fs. To vary the fundamental beam power we used two wire grid polarizers: the first one (rotatable) to control the incident THz power and the second one to keep the THz beam always vertically polarized.

*Simulations of grating enhancement* – The absorption and field enhancement in the TI-grating structure were calculated via rigorous coupled wave analysis (RCWA) with a 1D periodic pattern under TM polarization via the periodically patterned multilayer package [24, 28]. The geometry consisted of a 1 mm thick InP substrate, followed by the TI, capped with a 10 nm Al$_2$O$_3$ layer. On top of the Al$_2$O$_3$ there is the 50 nm thick gold grating, the only patterned part, with a fixed grating period of 33 μm, and the fill factor $F$ defined as the fraction of the period filled by the gold. Above and below the structure there are semi-infinite layers of air. The complex permittivities $\varepsilon = \varepsilon' + i\varepsilon''$ were taken as $\varepsilon_{Au} = -1.13 \cdot 10^5 + 3.48 \cdot 10^7 i$ for gold [29], $\varepsilon_{Al_2O_3} = 9.48 + 0.047i$ for Al$_2$O$_3$ [30], and $\varepsilon_{InP} = 12.5$ for InP [31]. We performed calculations considering the TI layer with a finite thickness (3D), as well as an interface only (2D), *i.e.* bulk vs. surface, respectively. In the case of the 3D implementation, the real part of the TI permittivity was taken as $\varepsilon' = 10, 180, 360$ (blue, orange, and green lines in Fig. 3b–c). We varied this because we found no consensus on this value in the literature. Figure 3 shows that this value does not play an important role in the final result. The imaginary part was calculated via $\varepsilon''_{TI} = \sigma_{3D}/(\varepsilon_0 \omega)$, and $\sigma_{3D} = e\mu_{TI} n_{TI}$, with the mobility $\mu_{TI} = 2000$ cm$^2$ V$^{-1}$ s$^{-1}$, carrier density $n_{TI} = 9 \cdot 10^{17}$ cm$^{-3}$ (from Ref. [25]), and frequency $\omega = 2\pi \cdot 0.5$ THz. Here, $\varepsilon_0$ is the vacuum permittivity and $e$ is the elementary charge. In the 2D implementation, the TI was modelled as a single conducting interface, with a 2D conductivity of $\sigma_{2D} = (1\ \text{k}\Omega)^{-1}$ (from Ref. [25]). In both cases, the resulting absorption was calculated for normal incidence and the enhancement factor was calculated with respect to the case of no gold grating.

*Calculation of surface-to-bulk "Coulomb cooling"* – The temperature dynamics of electrons in the surface state of a topological insulator is studied by considering their kinetic equation, i.e. $\partial_t f^{(s)}_{k,\lambda} = -\mathcal{I}^{(sb)}_{k,\lambda}$. The collision integral $\mathcal{I}^{(sb)}_{k,\lambda}$ describes the non-retarded and number-conserving Coulomb interaction between surface and bulk states:

$$\mathcal{I}^{(sb)}_{k,\lambda} = 2\frac{2\pi}{\hbar \mathcal{A}^2 L_z} \sum_{k',\tilde{k}',q} \sum_{\lambda'} \sum_{\eta,\eta'} \int_{-\infty}^{\infty} d\omega V^2_{q,k',\tilde{k}'} F^{(ss)}_{k,\lambda;k+q,\lambda'} \delta\left(\varepsilon^{(s)}_{k,\lambda} - \varepsilon^{(s)}_{k+q,\lambda'} + \omega\right) \delta\left(\varepsilon^{(b)}_{k',\eta} - \varepsilon^{(b)}_{\tilde{k}',\eta'} - \omega\right) \delta(k'_\parallel - \tilde{k}'_\parallel - q)$$
$$\times \left[f^{(s)}_{k,\lambda} f^{(b)}_{k',\eta}\left(1 - f^{(s)}_{k+q,\lambda'}\right)\left(1 - f^{(b)}_{\tilde{k}',\eta'}\right) - \left(1 - f^{(s)}_{k,\lambda}\right)\left(1 - f^{(b)}_{k',\eta}\right) f^{(s)}_{k+q,\lambda'} f^{(b)}_{\tilde{k}',\eta'}\right], \quad (1)$$

where the factor 2 upfront accounts for the spin degeneracy of bulk states, $\omega$ and $q$ are the transferred energy and momentum parallel to the surface, $k'_\parallel$ and $\tilde{k}'_\parallel$ are the components of the three-dimensional momenta $k'$ and $\tilde{k}'$ parallel to the surface, $\mathcal{A}$ is the surface area of the topological insulator and $L_z$ its extension in the third dimension. In Eq. (1), $F^{(ss)}_{k,\lambda;k+q,\lambda'}$ is the squared matrix element of the surface-electron density operator between incoming and outgoing scattering states. The matrix element of the screened Coulomb interaction $V_{q,k',\tilde{k}'}$ is obtained by (i) solving the electrostatic problem for surface electrons in proximity to bulk ones and (ii) by integrating the result over the incoming and outgoing bulk states of momenta $k'$ and $\tilde{k}'$. The surface and bulk electrons are described with two distinct Fermi distribution functions, $f^{(s)}_{k,\lambda}$ and $f^{(b)}_{k',\eta}$, characterized by chemical potentials $\mu_s$ and $\mu_b$, and temperatures $T_s$ and $T_b$, respectively. Here, $k'$ and $\tilde{k}'$ are the respective particle momenta, $\lambda = \pm 1$ denote the two surface bands forming a Dirac cone, while $\eta = \pm 1$ denote the bulk conduction and valence bands. The surface bands are approximated as $\varepsilon^{(s)}_{k,\lambda} = \lambda \hbar v_F |k|$, where $v_F$ is the surface Fermi velocity, while the bulk ones are $\varepsilon^{(b)}_{k',\eta} = \eta[\Delta + |k'|^2/(2m)]$. Here, $\Delta$ is half the bulk band gap and $m$ the bulk effective mass. For simplicity, we assume bulk bands to be spherically symmetric and having identical effective masses. Furthermore, we assume them to have no spin structure. We obtain the cooling time by multiplying the kinetic equation by the energy of the surface state, $\varepsilon^{(s)}_{k,\lambda}$, and summing over $k$ and $\lambda$. The equation then assumes the form

$$C_s \partial_t T_s = \frac{4}{\pi \hbar} \int \frac{d^2 q}{(2\pi)^2} V_q^2 \int_0^\infty d\omega \omega [n^{(s)}(\omega) - n^{(b)}(\omega)] \Im m \chi_b(q,\omega) \Im m \chi_s(q,\omega), \quad (2)$$

where $C_s$ is the heat capacity of surface states while $n^{s/b}(\omega) = \left[e^{\hbar\omega/k_B T_{s/b}} - 1\right]^{-1}$ are Bose-Einstein distributions at the surface and bulk temperatures, respectively. Finally, $\Im m \chi_\text{s}(\boldsymbol{q},\omega)$ and $\Im m \chi_\text{b}(\boldsymbol{q},\omega)$ are the imaginary parts of the density-density response functions of surface and bulk states, respectively, which are calculated with standard methods of linear-response theory [32]. These describe the absorption and emission spectra of the two systems, and therefore their product is proportional to the joint probability of a virtual (non-retarded) photon to be emitted by surface states and absorbed by bulk ones. The cooling time $\tau$ is finally obtained by recasting Eq. (2) in the form $C_\text{s} \partial_t T_\text{s} = -(T_\text{s} - T_\text{b})/\tau$. In the numerical calculations, we set the surface Fermi velocity $v_\text{F} = 0.5 \cdot 10^6$ m s$^{-1}$, the surface and bulk electron densities to $n_\text{s} = 10^{16}$ m$^{-2}$ and $n_\text{b} = 0$ (which implies that $\mu_\text{b} = 0$), respectively, $T_\text{b} = 300$ K, $m = 0.21 m_\text{e}$, where $m_\text{e}$ is the bare electron mass, $\Delta = 100$ meV, and the undoped-topological-insulator dielectric constant $\varepsilon_\text{b} = 10$.

**Data availability**

All data is available from the corresponding authors upon reasonable request. Supplementary information accompanies the manuscript on the *Light: Science & Applications* website (http://www.nature.com/lsa).

**Acknowledgements**

Parts of this research were carried out at ELBE at the Helmholtz Zentrum Dresden-Rossendorf e.V., a member of the Helmholtz Association. Simone Zanotto is acknowledged for developing the RCWA code dedicated to the treatment of the two-dimensional conducting interface. We acknowledge discussions with G. Kh. Kitaeva and K. A. Kuznetsov. K.J.T. acknowledges funding from the European Union's Horizon 2020 research and innovation program under Grant Agreement No. 804349 (ERC StG CUHL), RYC fellowship No. RYC-2017-22330, and IAE project PID2019-111673GB-I00. ICN2 was supported by the Severo Ochoa program from Spanish MINECO Grant No. SEV-2017-0706. A.P. acknowledges support from the European Commission under the EU Horizon 2020 MSCA-RISE-2019 programme (project 873028 HYDROTRONICS) and from the Leverhulme Trust under the grant RPG-2019-363. S.S., K.B., T.K., and L.W.M. acknowledge support by the SFB1170 (DFG project ID 258499086). C.G. and S.O.V. acknowledge support from the European Union's Horizon 2020 FET-PROACTIVE project TOCHA under grant agreement 824140. T.K., L.W.M., and G.V.A. acknowledge the financial support from the Würzburg-Dresden Cluster of Excellence on Complexity and Topology in Quantum Matter (EXC 2147, 39085490).

**Conflict of interests**

There are no conflicts of interest to declare.

**Contributions**

S.K., G.V.A., and K.J.T. initiated the project; S.S., K.B., and G.K. designed and fabricated the Bi$_2$Se$_3$ films, T.K. G.V.A. and L.W.M. coordinated all efforts on the Bi$_2$Se$_3$ films; C.G. and S.O.V. prepared the Bi$_2$Te$_3$ films; D.S.R. and S.V. performed the fabrication of metal gratings on the TI samples, supervised by K.J.T.; S.K., I.I., A.P., T.O., M.C., and J.C.D carried out the THz harmonics measurements.; A.P. developed the theoretical model for surface-state cooling dynamics with input from K.J.T.; A.B. performed the RCWA simulations, supervised by K.J.T.; K.J.T. wrote the manuscript with input from D.S.R., S.K., G.V.A., and all other authors.